\documentclass[a4paper,10pt,5p,preprint]{elsarticle}
\usepackage[utf8]{inputenc}
\usepackage[colorlinks=true]{hyperref}
\usepackage{microtype}
\usepackage{amsmath,bbm,amssymb}
\usepackage{xcolor,graphics,graphicx}
\usepackage[caption=false]{subfig}
\usepackage[export]{adjustbox}

\newcommand{\I}{i}
\newcommand{\Det}{\mathrm{Det}}
\newcommand{\T}{\mathsf{T}}
\newcommand{\Tr}{\mathrm{Tr}}
\newcommand{\e}{\mathrm{e}}
\newcommand{\be}{\begin{equation}}
\newcommand{\ee}{\end{equation}}
\newcommand{\bea}{\begin{align}}
\newcommand{\eea}{\end{eqnarray}}
\newcommand{\id}{\mathbbm{1}}

\newcommand{\sig}{{\boldsymbol{\sigma}}}

\newcommand{\rvec}{\boldsymbol{r}}
\newcommand{\bmat}{\begin{pmatrix}}
\newcommand{\emat}{\end{pmatrix}}

\biboptions{sort&compress}

\begin{document}

\title{Time-local optimal control for parameter estimation in the Gaussian regime}


\author[1]{Alexander Predko}
\author[2,3]{Francesco Albarelli}
\ead{francesco.albarelli@gmail.com}
\author[1]{Alessio Serafini}
\ead{serale@theory.phys.ucl.ac.uk}

\address[1]{Department of Physics \& Astronomy, University College London, Gower Street, WC1E 6BT, London, United Kingdom}
\address[2]{Faculty of Physics, University of Warsaw, 02-093 Warszawa, Poland}
\address[3]{Department of Physics, University of Warwick, Gibbet Hill Road, CV4 7AL, Coventry, United Kingdom}
	

\date{\today}

\begin{abstract}
Information about a classical parameter encoded in a quantum state can only decrease if the state undergoes a non-unitary evolution, arising from the interaction with an environment.
However, instantaneous control unitaries may be used to mitigate the decrease of information caused by an open dynamics.
A possible, locally optimal (in time) choice for such controls is the one that maximises the time-derivative of the quantum Fisher information (QFI) associated with a parameter encoded in an initial state.
In this study, we focus on a single bosonic mode subject to a Markovian, thermal master equation, and determine analytically the optimal time-local control of the QFI for its initial squeezing angle (optical phase) and strength.
We show that a single initial control operation is already optimal for such cases and quantitatively investigate  situations where the optimal control is applied after the open dynamical evolution has begun.
\end{abstract}


\maketitle

\section{Introduction}
In analysing the accuracy of high precision metrological setups, one must take into account quantum mechanics, since it enforces an intrinsically statistical description of experiments.
The field of quantum parameter estimation was born to address the challenges that arise when combining concepts from classical statistics with the formalism of quantum mechanics~\cite{helstrom1976quantum,Holevo2011b}.
This line of research has lead to the realisation that peculiar properties of quantum systems can be employed to build high precision sensors, with performances not available to purely classical systems~\cite{Giovannetti2011,Demkowicz-Dobrzanski2015a,Degen2016,Braun2018,Pezze2018}.

However, when the quantum system is subjected to a noisy (non-unitary) dynamics the promised advantage can easily be lost~\cite{Huelga1997,Escher2011,Demkowicz-Dobrzanski2012}.
Several approaches have been proposed to suppress or at least mitigate the effect of noise in quantum metrology.
Some of those assume (partial) access to the environment causing the non-unitary dynamics~\cite{Zheng2015a,Gefen2016,Plenio2016,Albarelli2018a,Ma2018,Albarelli2019a}, possibly also applying measurement-based feedback.
Another popular way of tackling this issue is the use of error correcting codes, which have to be specifically tailored for metrological applications~\cite{Arrad2014,Dur2014,Kessler2014,Lu2015,Layden2017,Zhou2017,Layden2019,Kapourniotis2019,Gorecki2019,Zhou2019e}.


The form of quantum control that is closest to experimental implementations is focused on devising optimal time-dependent pulses given an available set of control Hamiltonians~\cite{Koch2016}.
In this context, optimal pulses for noisy metrology have been investigated by numerical methods~\cite{Liu2017g,Liu2017d,Xu2019a}.
Dynamical decoupling~\cite{Viola1999,Suter2015} is a particularly useful approach in choosing pulses to protect a quantum system from an environment, and it has also been studied in the context of quantum metrology~\cite{Tan2014b,Sekatski2015a}.

We mention that the interplay between quantum metrology and quantum control is a vast subject that goes well beyond the aim of counteracting and mitigating the effect of noise.
Optimal control pulses have been employed to achieve a nonclassical time-scaling for the precision in the estimation of parameters characterizing time-dependent fields~\cite{Pang2017,Gefen2017a,Naghiloo2017}, and also to enhance the performance of remote parameter estimation~\cite{Kiukas2017a}.

In some physical systems, it is reasonable to assume that the control operations can act essentially instantaneously (relative to the timescales of the free dynamics).
Under this assumption, we can think of the controlled dynamics as the free nonunitary evolution interspersed by unitary operations; this is the framework we will be working in.
The use of such control unitaries was shown to be useful to restore a nonclassical time scaling when sensing nontrivial Hamiltonian parameters~\cite{Yuan2015,Yuan2016b,Hou2019}, and the same idea has been applied to Gaussian multimode interferometry~\cite{Matsubara2018}.

In this paper, we focus on a particular flavour of optimal control that is ``time-local', meaning that the time-derivative of some figure of merit is optimised over instantaneous unitary operations.
This approach has been applied to thermodynamical quantities, both in finite-dimensional~\cite{Mukherjee2013} and Gaussian quantum systems~\cite{Carlini2014,Shackerley-Bennett2017a} and also to counteract the decay of entanglement in two-mode Gaussian systems~\cite{Albarelli2018c}.
Here we follow the same approach but with the aim of counteracting the decay of the metrological usefulness of a quantum state, quantified by the quantum Fisher information (QFI).
In particular, we focus on a simple but paradigmatic model: a single bosonic mode evolving in a Markovian thermal environment.
Furthermore, we concentrate on Gaussian states and Gaussian control operations, since they are readily implementable in many physical platforms~\cite{serafini2017quantum,Ferraro2005,Weedbrook2012,Adesso2014}.

We consider a situation in which the encoding of the unknown parameter happens before the free open dynamics of the system.
This is a reasonable model when also the parameter encoding happens very quickly compared to the time scale of the free evolution.
In other words, the noisy dynamics can be considered as an unavoidable part of the detection stage, which does not affect the parameter encoding stage.
This approach is different from considering the encoding on the parameter as part of the dynamical evolution, as done in many of the studies we have mentioned.

One of our main results and our starting point for this analysis is a compact analytical formula for the time derivative of the QFI.
We find that the information about a parameter encoded in the first moments is unaffected by the control strategies we consider, and thus we restrict to parameters encoded in the covariance matrix (CM), i.e., angle and strength of squeezing.
Not surprisingly, the estimation of squeezing with Gaussian states has been considered by various authors~\cite{Milburn1994,Chiribella2006d,Gaiba2009,Genoni2009,Safranek2015,Safranek2016,Rigovacca2017}.
Here, we find the optimal unitary controls to preserve the QFI associated with both the squeezing angle (i.e., an optical phase parameter) and the squeezing strength.
Interestingly, the optimal strategy we find is analogous to the one found in~\cite{Yuan2015,Yuan2016b} for noiseless Hamiltonian estimation, i.e., applying the inverse of the parameter encoding unitary.
This inverse transformation also appears when studying optimal measurements for noiseless multiparameter estimation~\cite{Humphreys2013,Pezze2017}.
Since these optimal operations depend on the unknown true value of the parameter, their effectiveness only make sense in the context of an adaptive estimation scheme~\cite{Barndorff-Nielsen2000}.

Remarkably, we also find that repeated control operations are unnecessary and the optimal control operation is equivalent to an optimal encoding of the state before it undergoes the open dynamics.
We also study the effect of applying the optimal control operations with some delay after the start of the open evolution of the state.
Applying an initial control operation is equivalent to optimally encoding information before the action of a noisy channel, an idea that has been proposed to preserve the multipartite entanglement and the QFI of multiqubit systems~\cite{Chaves2012,Brask2015} and recently tested experimentally~\cite{Proietti2019}.
In a continuous variables setting, 
one may, along similar lines,
use squeezing to minimise the decoherence of non-Gaussian states evolving in a Gaussian environment, which is formally equivalent to squeezing the state of the environment~\cite{Serafini2004a,Serafini2005,Filip2013,Jeannic2017}.


This paper is structured as follows.
In Section~\ref{sec:system}, we introduce Gaussian systems and the dynamical model we will consider in the following.
In Section~\ref{sec:param}, we revise the basics of quantum parameter estimation, in particular applied to Gaussian systems. 
Section~\ref{sec:control} contains our results about time-local optimal control applied to the preservation of the QFI of Gaussian quantum states.
Section~\ref{sec:concl} ends the paper with some remarks and possible directions for future studies.


\section{System and dynamics}\label{sec:system}

We shall consider a system of one bosonic, mode associated to a vector of canonical operators $\hat{\boldsymbol{r}} = (\hat{x},\hat{p})^{\sf T}$ obeying the canonical commutation relations (CCR) $[\hat{x}, \hat{p}] = \I \hat{\id} $, where we have set 
$\hbar=1$. 
One may also express the CCR 
in terms of the symmetrised commutator \cite{serafini2017quantum} $[\hat{\boldsymbol{r}}, \hat{\boldsymbol{r}}^{\sf T}] = \hat{\boldsymbol{r}} \hat{\boldsymbol{r}}^{\sf T} - (\hat{\boldsymbol{r}}\hat{\boldsymbol{r}}^{\sf T})^{\sf T} =i\Omega$, where $\Omega$ is a 
$2  \times 2 $ matrix known as the symplectic form:
$$\Omega = \begin{pmatrix} 0 & 1 \\
                              -1 & 0\end{pmatrix}\; .$$
For a quantum state $\hat{\varrho}$, the expectation value of the observable $\hat{x}$ is given by $\langle \hat{x} \rangle =  \text{Tr}[\hat{\varrho} \hat{x}]$. 
Using  vector notation, this can be generalised to give the first and second statistical moments of a state:
\be \label{moments}
    \rvec = \text{Tr}[\hat{\varrho} \hat{\boldsymbol{r}}] \quad ,\quad
    \boldsymbol{\sigma} = \text{Tr}[\{(\hat{\boldsymbol{r}} - \rvec), (\hat{\boldsymbol{r}} - \rvec)^{\sf T}\}\hat{\varrho}] \, .
\ee
The above definition leads to a real, symmetric CM $\sig$, satisfying $\sig \geq \I \Omega$.

We will consider the encoding of classical information (a real parameter's value) in Gaussian states, which may in general be defined as the ground and thermal states of quadratic Hamiltonians. 
Such states are fully characterised by first and second statistical moments, as defined above.
Unitary operations which map Gaussian states into Gaussian states are those generated by a quadratic Hamiltonian.
The effect of such operations on the vector of operators is a symplectic transformation $\hat{\boldsymbol{r}} \rightarrow S \hat{\boldsymbol{r}}$ 
where $S$ is a $2 \times 2$ real matrix belonging to the real symplectic group $\mathrm{Sp}_{2,{\mathbbm R}}$, i.e.~$S\Omega S^{\sf T} = \Omega$. 
The corresponding effect on the CM of the system is the transformation $ \boldsymbol{\sigma} \xrightarrow{} S \boldsymbol{\sigma}S^{\sf T}$, while first moments are transformed according to $\rvec \mapsto S \rvec$.  
It will be convenient to parametrise single-mode symplectic transformations though their singular value decomposition \cite{serafini2017quantum}:
\be
S =  \left(\begin{array}{cc}\cos\varphi &\sin\varphi\\
-\sin\varphi&\cos\varphi \end{array}\right) 
\left(\begin{array}{cc}z & 0 \\
0& \frac1z \end{array}\right)
\left(\begin{array}{cc}\cos\chi &\sin\chi\\
-\sin\chi&\cos\chi \end{array}\right)\, , \label{svd}
\ee
for $\chi,\varphi\in [0,2\pi[$ and $z\ge 1$.
From the normal mode decomposition of $\sig$ and Eq.~(\ref{svd}), it follows that the most general CM of a single-mode system 
may be written as a rotated and squeezed thermal state of the free Hamiltonian $(\hat{x}^2+\hat{p}^2)$ \cite{serafini2017quantum}:
\be
\sig = \nu  \left(\begin{array}{cc}\cos\theta &\sin\theta\\
-\sin\theta&\cos\theta \end{array}\right)
\left(\begin{array}{cc}y^2 &0\\
0&\frac{1}{y^2} \end{array}\right)
\left(\begin{array}{cc}\cos\theta &-\sin\theta\\
\sin\theta&\cos\theta \end{array}\right) \, ,\label{1m}
\ee
with squeezing parameter $y\ge 1$, optical phase $\theta\in[0,2\pi[$, we will also use the terms squeezing strength for the parameter $r$, defined as $y=\e^{r}$, and squeezing angle $\theta$.
These parameters appear in the unitary transformation $\hat{S}_{\xi} = \e^{ \frac{1}{4} \xi^*\hat{a}^2 - \frac{1}{4} \xi \hat{a}^{\dag 2}}$, with the complex parameter $\xi = r \e^{\I \theta}$, where we have introduced the bosonic annihilation operator $\hat{a} = (\hat{x} + \I \, \hat{p})/\sqrt{2}$.
The symplectic eigenvalue $\nu\ge 1$ captures the temperature and purity of the state; states with $\nu=1$ are pure, while the ground state of the free Hamiltonian (the ``vacuum'' state) is obtained by setting 
$y=\nu=1$, so that $\sig=\id$.

The free, uncontrolled dynamics of our system will be the diffusion induced by contact with a white noise (Markovian) environment at finite temperature, described by a Lindblad master equation
\begin{equation}
\dot{\hat{\varrho}} = \mathcal{L}_{\bar{N}} \hat{\varrho} = \left( \bar{N} + 1\right) \mathcal{D} [ \hat{a} ]  \hat{\varrho} + \bar{N} \mathcal{D} [ \hat{a}_i^\dag ]  \hat{\varrho}\, , \label{eq:LindbladTherm}
\end{equation}
where we have introduced the superoperator $\mathcal{D} \left[ \hat{o} \right] \! \hat{\varrho} = \hat{o} \hat{\rho} \hat{o}^\dag - \left\{ \hat{o}^\dag \hat{o} , \hat{\rho} \right\} $ and the Lindbladian $\mathcal{L}_{\bar{N}}$, i.e. the generator of the dynamical semigroup~\cite{Gorini1976,Lindblad1976}.
We work with dimension-less time, expressed in units of the inverse loss rate, that thus does not appear in~\eqref{eq:LindbladTherm}; $\Bar{N}$ is the mean number of excitations in the environment, related to its inverse temperature $\beta$ by the Bose law.
This dynamics describes loss in a thermal environment and is ubiquitous in quantum optics.
Since the generator is time-independent, the solution from a time $t_0$ to a time $t$ is formally given by the map $\e^{(t-t_0) \mathcal{L}_{\bar{N}}}$.
This evolution is also known as the quantum attenuator channel.

At the level of Gaussian states this dynamics is described by the following equations of motion for first and second moments:
\begin{align}
\dot{\sig} &= -\sig + N\id \; , \label{difsig}\\
\dot{\rvec} & = -\frac{\rvec}{2} \; \label{difr} ,
\end{align}
where $N=(2\Bar{N}+1)$.
For a more general and detailed description of Gaussian quantum systems and dynamics similar to the one presented here see~\cite{Genoni2016,serafini2017quantum}.

We shall assume the ability to intersperse such an open dynamics with instantaneous Gaussian (IG) unitary operations $\hat{U}_\mathsf{IG}$ that will enact our control, which correspond to symplectic transformations in the phase-space description suited to Gaussian states.
In practice, this is a reasonable assumption if the inverse loss rate is much larger than the time it takes to perform a control operation. 
Albeit such times may vary widely in specific cases, there exist optical set-ups with typical decoherence times around $10$ $\mu{\rm s}$ 
and operating times around $1$ ${\rm ns}$, where the hypothesis of instantaneous controls is very reasonable.

\section{Quantum parameter estimation}\label{sec:param}
Let $\hat{\varrho}_\theta$ be a set of quantum states whose exact form depends on an unknown parameter $\theta$.
The problem of the optimal estimation of $\theta$ (i.e., of obtaining an estimator with minimal variance)
through a fixed POVM $\boldsymbol{\Pi} = \{\hat{\Pi}_x \geq 0 \, , \,  \sum_x \hat{\Pi}_x = \id \}$,
is a classical problem associated with the probability distribution $p(x|\theta)=\mathrm{Tr}[\hat{\Pi}_x \hat{\varrho}_\theta ]$.
The optimal solution is given by the classical Fisher information $I_{\boldsymbol{\Pi},\theta}$
(where we emphasise the dependence on the chosen POVM in the context pictured above), 
given by
\be
I_{\boldsymbol{\Pi},\theta} = \sum_{\mu} p(\mu|\theta) \left[\partial_{\theta}\ln\left(p(\mu|\theta)\right)\right]^2 =
 \sum_{\mu} \frac{\left(p'(\mu|\theta)\right)^2}{p(\mu|\theta)} \; 
\ee
(where the prime denotes the partial derivative with respect to the parameter $\theta$), and 
by the associated Cram{\'e}r--Rao bound, which may be saturated by unbiased estimators:
\be
\Delta\theta \ge \frac{1}{\sqrt{n I_{\boldsymbol{\Pi},\theta}}} \; , \label{cr}
\ee
where $\Delta\theta$ is the standard deviation on the estimate of $\theta$ and $n$ is the number of measurements carried out (in the language of statistics this corresponds to the sample size).
We stress that, while an unbiased estimator might not exist for finite $n$, it is usually possible to find an estimator that saturates~\eqref{cr} in the limit of large $n$~\cite{Braunstein1992a,lehmann_theory_1998}.

The optimisation of the classical Fisher information over all possible POVMs gives rise to the QFI $I_{\theta}$~\cite{helstrom1976quantum,Holevo2011b,Braunstein1994,Paris2009}:
\be
I_{\theta} = \max_{\boldsymbol{\Pi}} I_{\boldsymbol{\Pi},\theta} = \lim_{\epsilon \rightarrow 0} \frac{ 8 \left(1 - F \left[ \hat{\varrho}_\theta, \hat{\varrho}_{\theta + \epsilon}  \right] \right) }{ \epsilon^2 }  \; , \label{qfidef}
\ee
where, as above, the symbol $\boldsymbol{\Pi}$ stands for the whole set of POVMs and the QFI is expressed in terms of the fidelity $F \left[ \varrho , \sigma  \right] = \left\Vert \sqrt{\varrho} \sqrt{\sigma} \right\Vert_1$, where $\Vert A \Vert_1 = \mathrm{Tr} \left[ \sqrt{A A^\dag} \right]$ is the trace norm.
One can prove that the QFI is monotonically decreasing when parameter independent channels are applied to the state, while it is invariant if the transformation is a unitary~\cite{Petz1996a,Hayashi2017c}.
As one would expect, the QFI enters the quantum Cram{\'e}r--Rao bound
\be
\Delta\theta \ge \frac{1}{\sqrt{ n I_{\theta}}} \; . \label{qcr}
\ee
Since this inequality may be shown to be achievable, it represents the ultimate bound to quantum parameter estimation.
The QFI is therefore the most fundamental quantity in assessing the sensitivity of a system to a certain parameter, which might reflect environmental or technical factors.
Just as in the classical case, the inequality~\eqref{qcr} is usually saturated only in the asymptotic limit.
However the optimal POVM $\mathrm{arg \,max}_{\boldsymbol{\Pi}} I_{\boldsymbol{\Pi},\theta}$ might depend on the true unknown value $\theta$, thus some kind of adaptive strategy is needed in general~\cite{Barndorff-Nielsen2000}.
In other words, the information quantified by the QFI makes sense in a \emph{local estimation} scenario, where we assume to have prior knowledge about the parameter value and to be in a neighbourhood of the true value of the parameter.

If the quantum states $\hat{\varrho}_{\theta}$ are all Gaussian states, and therefore the dependence on $\theta$ is entirely contained in the CM $\sig$ and in the vector of first moments $\rvec$, one may obtain an analytical formula 
for the QFI.
Such a formula is particularly wieldy for single-mode Gaussian states~\cite{Pinel2013,Monras2013,Jiang2014,Banchi2015,serafini2017quantum}, 
on which we shall focus in this paper:
\be
I_{\theta} = \frac12\frac{{\rm Tr}[(\sig^{-1}\sig')^2]}{1+\mu^2} +\frac{2\mu^{\prime 2}}{1-\mu^4}
+ 2{\rvec}^{\prime \sf T}\sig^{-1}\rvec' \; ,
\label{qfi}
\ee
where $\mu=\mathrm{Tr}[\hat{\varrho}_{\theta}^2]=1/\sqrt{\mathrm{Det}\sig}=1/\nu$ is the purity of the quantum states and the prime $'$ denotes differentiation with respect to the parameter $\theta$.
Notice that, in all of the formulae above, the derivatives are taken at the `true' value $\Bar{\theta}$ of the parameter $\theta$,
so that both the QFI and the optimal POVM will in general depend on $\Bar{\theta}$ (more on this issue later).
In the following, it will be convenient to make the distinction between $\theta$ and $\Bar{\theta}$ explicit and clear.

\section{Locally optimal control to protect the quantum Fisher information}
\label{sec:control}

Later on, we shall assume Gaussian states depending on an unknown optical phase or squeezing parameter and subject to the diffusive dynamics (\ref{difsig},\ref{difr}), and assess the performance of instantaneous control symplectics towards the task of parameter estimation.
We would like therefore to determine controls that maximise the evolving QFI associated with phase estimation.
Notice that, although the QFI is obviously invariant under unitary, and hence symplectic, operations, its time-derivative under the free open dynamics we are considering need not be, and in fact is not.
Therefore, symplectic controls may enhance the QFI during the time evolution.

Our first step is then to obtain a general expression for the time-derivative of the QFI (\ref{qfi}) under Eqs.~\eqref{difsig} and~\eqref{difr}, which 
is derived in~\ref{app:deriv} and turns out to be rather compact:
\begin{align}
& \dot{I_{\theta}} = \frac{\mu^2{\rm Tr}[(\sig^{-1}\sig')^2](N{\rm Tr}[\sig^{-1}]-2)}{2(1+\mu^2)^2} 
-\frac{N{\rm Tr}[(\sig^{-1}\sig')^2\sig^{-1}]}{1+\mu^2} \nonumber\\
&- \frac{\mu^2 {\rm Tr}[\sig^{-1}\sig']\left((N{\rm Tr}[\sig^{-1}]-2){\rm Tr}[\sig^{-1}\sig'] +2N {\rm Tr}[\sig^{-2}\sig'] \right)}{2(1-\mu^4)} \nonumber \\
& +\frac{\mu^6}{(1-\mu^4)^2}{\rm Tr}[\sig^{-1}\sig']^2(2-N{\rm Tr}[\sig^{-1}]) - 2N{\bf r}'^{\sf T} \sig^{-2}{\bf r}' \; . \label{idiot}
\end{align}
The ``locally'' (in time) optimal control will be determined by letting $\sig\mapsto S \sig S^{\sf T}$ (and likewise for $\sig'$) and 
${\bf r}\mapsto S\mathbf{r}$ in the expression above, and by maximising it with respect to the parameters $\chi$, $\varphi$ and $z$ that 
parametrise the control $S$ as per Eq.~(\ref{svd}).

Inspection reveals that the controlled $\dot{I_{\theta}}$ does not depend on the first rotation in the singular value decomposition of $S$, 
so that we can set $\chi=0$ in what follows without loss of generality.

Also note that the control can never help preserving the QFI associated with first moments, since the only term which depends on 
them is clearly invariant under any $S$. We will therefore neglect first moments hereafter.

\subsection{Estimation of the squeezing angle}

As a first case study, let us consider the estimation of the optical phase $\theta$ of a squeezed state with first moments independent from $\theta$ and CM of the form of Eq.~\eqref{1m}. 
Notice that this also encompasses the case of a state undergoing 
the dynamics described by Eqs.~(\ref{difsig},\ref{difr}) for any initial transient time, since the dynamics is phase-covariant, i.e. it commutes with the action of the phase shifter imprinting the dependence on the parameter $\theta$.
In other words, the only difference between considering the instantaneous control acting at the beginning or at some intermediate time is reflected in different values for the parameters $\nu$ and $y$, but the optimization problem to be solved remains identical.

It is now convenient to define 
\be
R_{\theta} = \left(\begin{array}{cc}\cos\theta &\sin\theta\\
-\sin\theta&\cos\theta \end{array}\right)
\ee
and observe that
\begin{align}
R_{\theta}' &= R_{\Bar{\theta}} \Omega = \Omega R_{\Bar{\theta}} \; , \\
R_{\theta}^{\sf T \prime} &= R_{-\theta}' = - R_{-\Bar{\theta}} \Omega = -\Omega R_{-\Bar{\theta}} \; 
\end{align}
(recall that all the derivatives with respect to the parameter $\theta$ must be taken at the true value $\Bar{\theta}$),
where $\Omega$ is the $2\times 2$ symplectic form.
Let us also define the diagonal squeezing matrices $Y^{2}={\rm diag}(y^2,1/y^2)$, which characterises the initial state, and 
$Z={\rm diag}(z,1/z)$, characterising the control operation.

Including the transformation (\ref{svd}) for $\chi=0$, one has the following expressions for the controlled $\sig^{-1}$ and $\sig'$
associated to the estimation of the phase of squeezing:
\begin{align}
\sig^{-1} &= \nu^{-1} Z^{-1} R_{(\Bar{\theta}+\varphi)} Y^{-2} R_{-(\Bar{\theta}+\varphi)} Z^{-1} \; , \\
\sig' &= \nu Z R_{(\Bar{\theta}-\varphi)} [\Omega,Y^2] R_{(\varphi-\Bar{\theta})} Z \nonumber \\ &= 
\nu \left(\frac{1}{y^2}-y^2\right)
Z R_{(\Bar{\theta}-\varphi)} \sigma_x R_{(\varphi-\Bar{\theta})} Z \, ,
\end{align}
where $\sigma_x$ is the Pauli $x$ matrix.
Whence
\be
\sig^{-1}\sig' =  \left(\frac{1}{y^2}-y^2\right) Z^{-1} R_{(\Bar{\theta}+\varphi)} Y^{-2} \sigma_x R_{-(\Bar{\theta}+\varphi)} Z
\ee
and therefore, in this case,
\be
{\rm Tr}[\sig^{-1}\sig'] = \nu^{-1} \left(\frac{1}{y^2}-y^2\right) {\rm Tr}[Y^{-2}\sigma_x] = 0 \,,
\ee
which simplifies our task greatly, since it sets to zero the third and fourth term in Eq.~(\ref{idiot}) which, noticing 
that ${\rm Tr}[(\sig^{-1}\sig')^2]=  2\left(\frac{1}{y^2}-y^2\right)$ and ${\rm Tr}[(\sig^{-1}\sig')^2\sig^{-1}]=
\left(\frac{1}{y^2}-y^2\right) {\rm Tr}[\sig^{-1}]$, reduces to:
\be
\dot{I_{\theta}} = -\left(\frac{1}{y^2}-y^2\right) \frac{2\nu^2 + N\nu^2(2\nu^2 + 1){\rm Tr}[\sig^{-1}]}{2(\nu^2+1)^2} \; .
\ee
The optimal control operation is therefore the one that minimises ${\rm Tr}[\sig^{-1}]$, whose coefficient above is negative.
It is apparent from the general expression of a single-mode CM~\eqref{1m} that, up to the symplectic eigenvalue that is not affected by a symplectic transformation, ${\rm Tr}\sig^{-1}$ is minimised, obtaining the value $2/\nu$, 
by making the CM proportional to the identity, which can always be done through a symplectic control that undoes the initial rotation $\Bar{\theta}$ and squeezing $z$, i.e. by choosing $\varphi = - \Bar{\theta}$ and $z=1/y$.

Notice that such an optimal control depends on the unknown value $\Bar{\theta}$.
This is however not too worrisome in practice, as such a value may be estimated with a first run of measurements without controls, which would provide the hypothetical experimentalist with an approximation of the optimal transformation.
This situation is common for \emph{local} quantum estimation theory and it is analogous to the well known fact that the optimal POVM, attaining the quantum Cram{\'e}r--Rao bound, also depends on $\Bar{\theta}$~\cite{Barndorff-Nielsen2000}.

Moreover, very remarkably, once the initial squeezing is undone and the state is brought in the optimal form, no further control is required to maintain optimality, since the free diffusive dynamics does not change the form of the (unsqueezed) CM: a single manipulation is therefore optimal among all possible control strategies.


\begin{figure}
\includegraphics[width=\columnwidth]{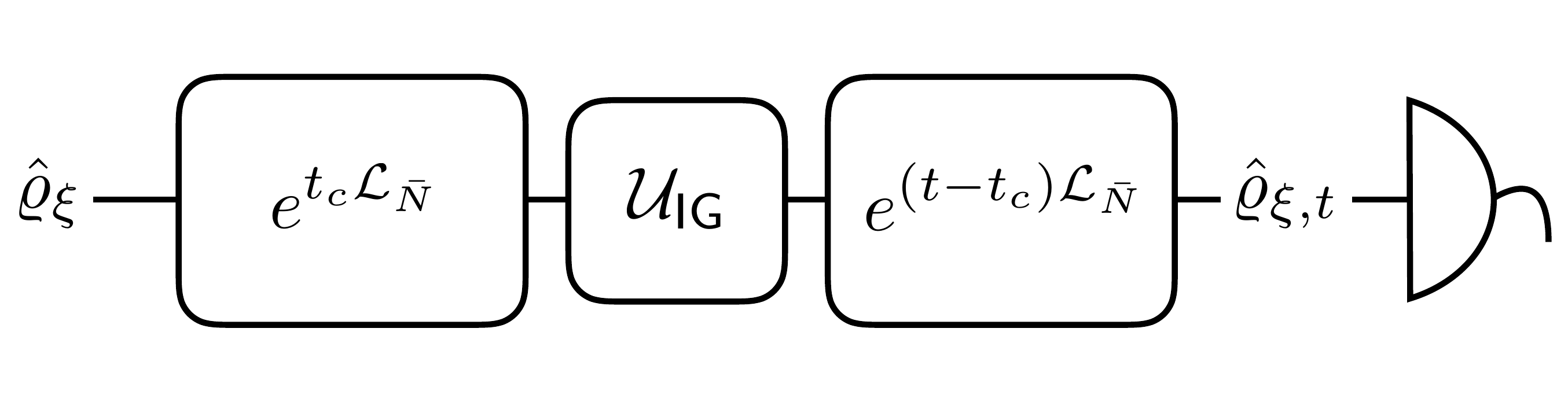}
\caption{Schematic representation of the control protocol we consider.
A squeezed vacuum state $\hat{\varrho}_\xi = \hat{S}_\xi |0 \rangle \langle 0 |\hat{S}_\xi^\dag$ is subjected to the open dynamics describing loss in a thermal environment and obeys the master equation~\eqref{eq:LindbladTherm}, or equivalently equation~\eqref{difsig} for the CM, since the state has zero first moments.
At time $t_c$ (in units of the inverse loss rate) a control operation is applied to the system, in the form of a instantaneous Gaussian unitary $\mathcal{U}_{\mathsf{IG}} \hat{\varrho} = \hat{U}_\mathsf{IG} \hat{\varrho} \hat{U}_\mathsf{IG}^\dag$, with the aim of slowing down the decrease in metrological usefulness of the state, quantified by the QFI.
The evolution then proceeds to the final time $t$ and the evolved state $\hat{\varrho}_{\xi,t}$ is measured.
The metrological task is to estimate the value of either $\phi$ (squeezing angle) or $r$ (squeezing strength), where $\xi = r \e^{i \phi}$.}
\label{fig:scheme}
\end{figure}

Summing up, we have shown that the sensitivity of squeezed states to their optical phase is enhanced if the squeezing is undone through a single control operation.
In order to quantify the advantage granted by optimal instantaneous symplectic controls, it is expedient to assume that, at the time $t=0$, an initial pure squeezed vacuum state $\hat{\rho}_\xi = \hat{S}_{\xi} |0\rangle \langle 0 | \hat{S}_{\xi}^\dag$, with any given $\Bar{\theta}$ and $y$ and $\nu=1$, starts being subject to the diffusive dynamics \eqref{difsig} and then undergoes optimal control at a time $t_c\ge 0$.
This setup is schematically represented in Fig.~\ref{fig:scheme}.

Before the control is applied (i.e., for $t<t_c$), one has  
\begin{align}
\sig &= {\rm e}^{-t} R_{\Bar{\theta}} Y^2 R_{-\Bar{\theta}} + (1-{\rm e}^{-t})N\id \; , \\
\sig'&= {\rm e}^{-t} \left(\frac{1}{y^2}-y^2\right) R_{\Bar{\theta}} \sigma_x R_{-\Bar{\theta}} \; ,
\end{align}
which may be inserted into Eq.~(\ref{qfi}) to obtain the QFI of the evolved state (as shown in~\ref{app:deriv}, the second term on the RHS of Eq.~(\ref{qfi}) is proportional 
to ${\rm Tr}[\sig^{-1}\sig']$ and hence vanishes in our case, and so does the third as the first moments do not contribute to the estimation).

After the control is applied, for $t\ge t_c$, one has instead
\begin{align}
\sig &= {\rm e}^{-t} Y_{c}^{-1} R_{(\Bar{\theta}-\varphi_c)} Y^2 R_{-(\Bar{\theta}-\varphi_c)} Y_{c}^{-1} +(1-{\rm e}^{-t})N Y_{c}^2  
\nonumber \\
 &= \left({\rm e}^{-(t-t_c)}\nu_c +(1-{\rm e}^{-(t-t_c)})N \right) \id \; , \\
\sig'& = {\rm e}^{-t} 
{\rm e}^{-t} \left(\frac{1}{y^2}-y^2\right) Y_{c}^{-1} \sigma_x Y_{c}^{-1} =  
{\rm e}^{-t} \left(\frac{1}{y^2}-y^2\right) \sigma_x \; , 
\end{align}
where $\varphi_c=\Bar{\theta}$ (the distinction between the two has been maintained to make the derivation of 
$\sig'$ clearer), the diagonal squeezing transformation $Y_{c}$ is chosen {\em ad hoc} to make the CM 
proportional to the identity, and 
$\nu_c = \sqrt{\Big[{\rm e}^{-t_c}y^2+(1-{\rm e}^{-t_c})N\Big]
\Big[\frac{{\rm e}^{-t_c}}{y^2}+(1-{\rm e}^{-t_c})N\Big]}$ is the symplectic eigenvalue of the evolving state at the moment the control is enacted.

\begin{figure}[t!]
\includegraphics[width=.47\textwidth,left]{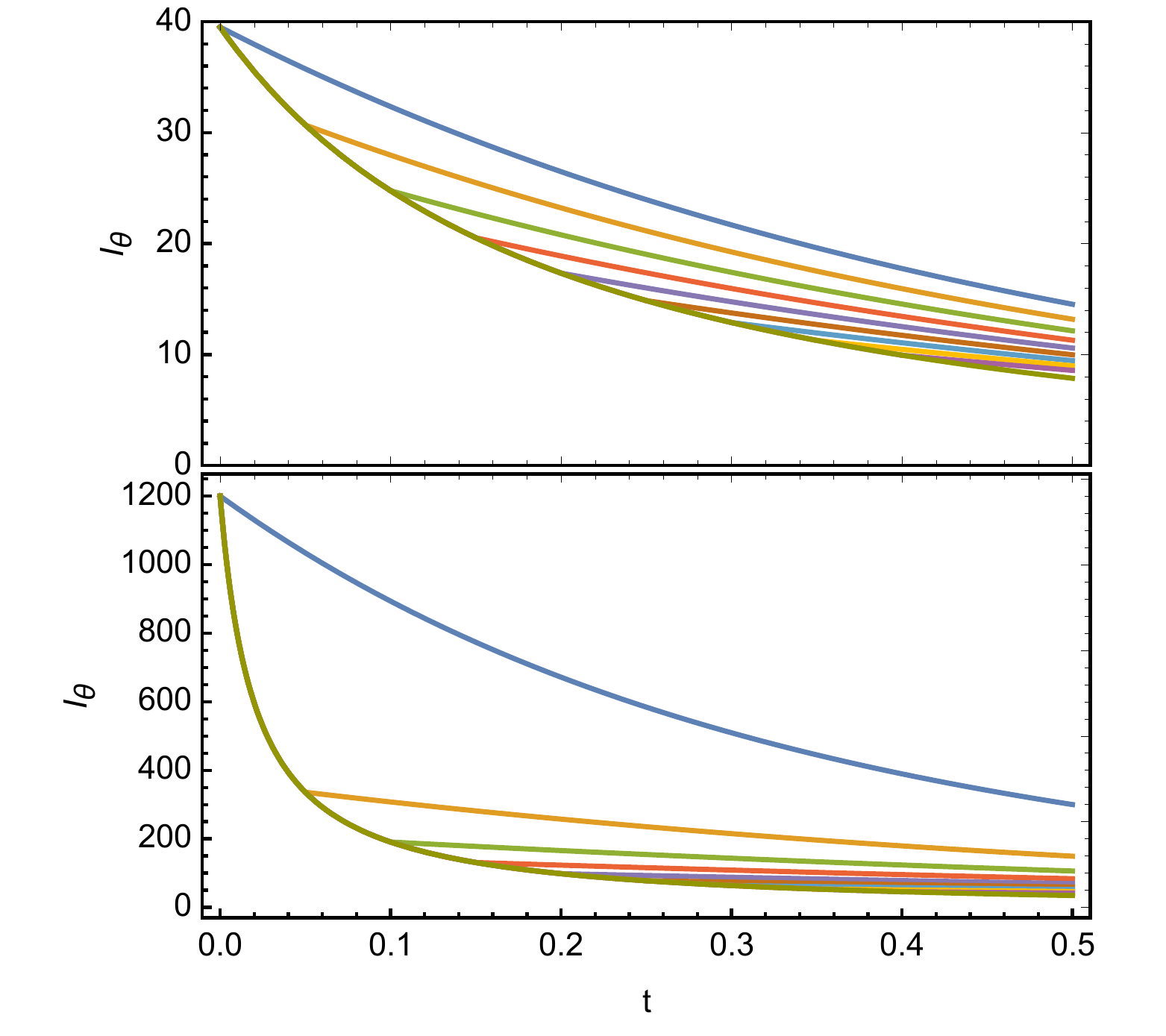}
\caption{
The QFI for the estimation of the squeezing angle $\theta$ as a function of time (both dimensionless, the latter in units of inverse loss rate), shown for different times at which the locally optimal control is applied:
the top (blue) curve depicts the case where the control is applied at $t=0$; at each subsequent curve, from top to bottom, the control is applied $0.05$ inverse loss rates later; the bottom curve (green) depicts the uncontrolled QFI. 
In the upper panel we consider a system with initial squeezing $y=3$ and an open dynamics with $N=1$ (``pure loss''), while in the lower panel we consider $y=10$ and $N=2$.
\label{Phestering}}
\end{figure}
The calculations detailed in the two previous paragraphs yield the following expression for the ``controlled'' QFI as a function of time:
\be
I_{\theta} = \left\{\begin{array}{ll} \frac{{\rm e}^{-2t} \left(\frac{1}{y^2}-y^2\right)^2}{\Big[{\rm e}^{-t}y^2+(1-{\rm e}^{-t})N\Big] \Big[\frac{{\rm e}^{-t}}{y^2}+(1-{\rm e}^{-t})N\Big]+1} \;  , &\; t<t_c\, , \\
\\
\frac{{\rm e}^{-2t} \left(\frac{1}{y^2}-y^2\right)^2}{ \left({\rm e}^{-(t-t_c)} \nu_c + (1-{\rm e}^{-(t-t_c)})N\right)^2 +1} \; , &\; t\ge t_c \, . 
\end{array} \right.
\ee
The effect of such a control scheme on the QFI is illustrated in Fig.~\ref{Phestering}, where one may appreciate the advantage gained by activating the control protocol after different intervals from the initial time.
By comparing the two panels we can also appreciate on a qualitative level that an initial control operation is more useful when the initial squeezing is high, since in this case the uncontrolled QFI drops more steeply.

\subsection{Estimation of the squeezing strength}

The same analysis carried out above for the squeezing phase may be repeated for the squeezing strength of the initial state.
To this aim, it is convenient to re-parametrise the state, setting $y^2 = \e^{r}$, so that $Y^{2}={\rm e}^{r \sigma_z}$ 
($\sigma_z$ being the Pauli $z$ matrix), and consider the estimation of the parameter $r$.
The problem becomes then completely phase-invariant, so we can omit the rotation in the evolving state without 
loss of generality, and just assume the CM $\nu {\rm e}^{r \sigma_z}$, for $r>0$ and $\nu\ge 1$. 

The controlled $\sig^{-1}$ and $\sig^{-1}$
associated to the estimation of the squeezing parameter $r$ take the form
\begin{align}
\sig^{-1} &= \nu^{-1} Z R_{-\varphi} {\rm e}^{-r\sigma_z} R_{\varphi} Z \; , \\
\sig' &= \nu Z^{-1} R_{-\varphi} {\rm e}^{r\sigma_z} \sigma_z  R_{\varphi} Z^{-1}  \; .
\end{align}
Whence
\be
\sig^{-1}\sig' =   Z R_{-\varphi} \sigma_z R_{\varphi} Z^{-1} \; ,
\ee
such that, once again, 
${\rm Tr}[\sig^{-1}\sig'] = 0$, and ${\rm Tr}[(\sig^{-1}\sig')^2]=  2$, ${\rm Tr}[(\sig^{-1}\sig')^2\sig^{-1}]=
 {\rm Tr}[\sig^{-1}]$, leading to
\be
\dot{I_r} = - \frac{2\nu^2 + N\nu^2(2\nu^2 + 1){\rm Tr}[\sig^{-1}]}{2(\nu^2+1)^2} \; .
\ee
In this case too, the optimal control operation is the one that minimises ${\rm Tr}[\sig^{-1}]$
that, as above, is obtained by setting $Z = {\rm e}^{r\sigma_z}$ and $\varphi=0$.
The optimal strategy requires a single control, since the free dynamics preserve the optimal form of the CM (of minimal ${\rm Tr}[\sig^{-1}]$ upon symplectic action).

\begin{figure}[t!]
\includegraphics[width=.47\textwidth,left]{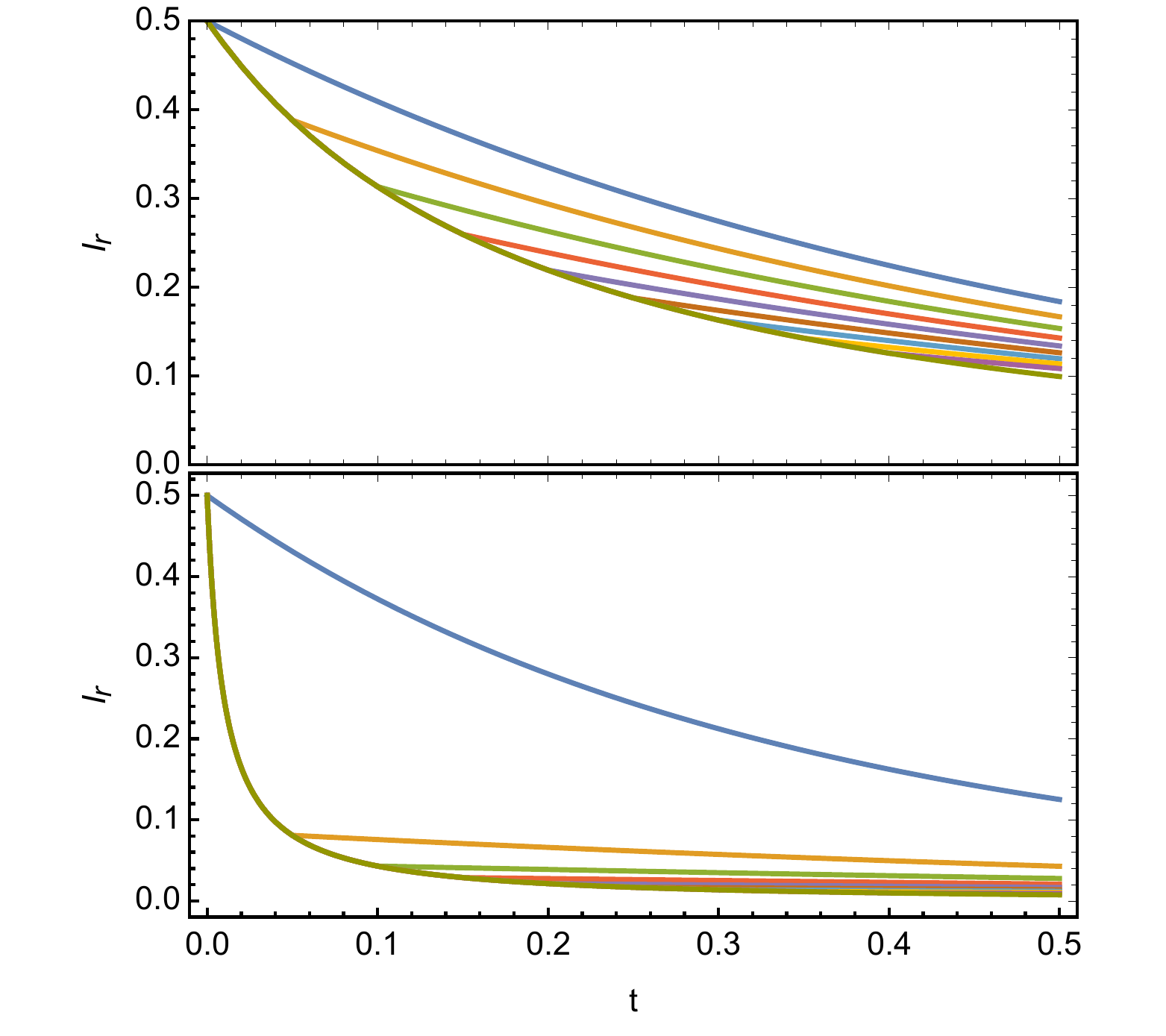}
\caption{The QFI for the estimation of the squeezing strength as a function of time (both dimensionless, the latter in units of inverse loss rate), shown for different times at which the locally optimal control is applied:
the top (blue) curve depicts the case where the control is applied at $t=0$; at each subsequent curve, from top to bottom, the control is applied $0.05$ inverse loss rates later; the bottom curve (green) depicts the uncontrolled QFI. 
In the upper panel we consider a system with initial squeezing $y=3$ and an open dynamics with $N=1$ (``pure loss''), while in the lower panel we consider $y=10$ and $N=2$.
\label{Phestering2}}
\end{figure}

Notice that the formulae for the estimation of the squeezing parameter are identical to what found above, except for the absence of the factor $(y^{-2}-y^2)$ in $\sig'$ (clearly, for the optical phase, the QFI is always equal to $0$ for $y=1$, because the initial state is then rotationally invariant; this is not an issue for squeezing, since such an operation admits no invariant states). 
Assuming, like above, to start with a pure squeezed vacuum state $\hat{\rho}_\xi = \hat{S}_{\xi} |0\rangle \langle 0 | \hat{S}_{\xi}^\dag$, and that the control is applied at some time $t_c\ge 0$, one has the rescaled formula for the QFI as a function of time:
\be
I_r = \left\{\begin{array}{ll} \frac{{\rm e}^{-2t}}{\big[{\rm e}^{-t}y^2+(1-{\rm e}^{-t})N\big]
\big[\frac{{\rm e}^{-t}}{y^2}+(1-{\rm e}^{-t})N\big]+1} \; , &\; t<t_c\, , \\ 
\\
\frac{{\rm e}^{-2t}}{ \big[{\rm e}^{-(t-t_c)} \nu_c + (1-{\rm e}^{-(t-t_c)})N\big]^2+1} \; , &\; t\ge t_c \, . 
\end{array} \right.
\label{eq:cQFIr}
\ee
with $\nu_c^2 = \big[{\rm e}^{-t_c}y^2+(1-{\rm e}^{-t_c})N\big]
\big[\frac{{\rm e}^{-t_c}}{y^2}+(1-{\rm e}^{-t_c})N\big]$, exactly as in the previous section.
We plot this quantity in Fig.~\ref{Phestering2}: the behaviour in time is qualitatively the same as the previous case of phase estimation, albeit the scale is different because of the absence of the factor $(y^{-2}-y^2)^2$ in the expression for $I_r$.
In this case the initial QFI at $t=0$ is independent from the parameter and equal to $1/2$.

While the phase estimation considered in the previous section is not possible with an unsqueezed state, the strength can instead be estimated at the true value $r=0$.
However, as one would expect, in this particular case there is no need for control operations since the state is already thermal; this can also be checked explicitly from~\eqref{eq:cQFIr} by setting $y=1$.

\section{Conclusions and remarks}\label{sec:concl}
We have only started to uncover the usefulness of time-local quantum control for the task of parameter estimation.
In particular, Gaussian systems proved to be a useful test bed for these strategies, since their simplicity allowed us to derive closed form expressions for the relevant quantities at play.

Essentially, we have shown that, for the considered open dynamics, the locally-optimal (in time) way to delay the decay of QFI about a parameter unitarily encoded in the CM of a single-mode Gaussian state is to unsqueeze it, thereby transforming it into a thermal state.
This means that such a thermal state is best suited to withstand a lossy evolution in a thermal environment, which is mathematically described by a phase-covariant channel.
Intriguingly, our results might be related to the fact that the minimum output entropy of a phase-covariant Gaussian channel is achieved by a thermal input state~\cite{DePalma2017g,DePalma2018}, although we lack a deeper understanding of this connection.

Several extensions of the ideas we have proposed here can be envisioned, most notably dropping the assumption of fast parameter encoding.
This means that, instead of only delaying the demise of the QFI of an initial state, we should consider the situation where the parameter is encoded simultaneously to the open dynamics, e.g., by adding an Hamiltonian term in the Lindblad master equation or estimating parameters of the non-unitary part of the Gaussian dynamics, see, e.g.,~\cite{Monras2007,Genoni2016a,Rossi2016,McMillen2017a,Nichols2018,Branford2019,Binder2019}.
In such scenarios time-local control would be used to increase the rate at which information about the parameter is acquired during the dynamics.

Finally, let us briefly mention that there exist dynamical decoupling schemes tailored for continuous variable systems~\cite{Vitali1999,Arenz2017a} that could in principle be used to completely remove the effect of the environment.
However, in dynamical decoupling the control operations must be not only instantaneous, but they have to be applied with a rate greater than the environment cut-off frequency and are thus very hard to implement in practice.
On the other hand our proposed strategy, while only capable of mitigating the effect of noise, can be readily implemented on experimental platforms through a single control operation.

\section*{Acknowledgements}
\noindent We are grateful to D.~Branford, A.~Datta and M.~G.~Genoni for useful discussions.
FA acknowledges financial support from the UK National Quantum Technologies Programme (EP/M013243/1) and from the National Science Center (Poland) grant No. 2016/22/E/ST2/00559.

\appendix
\section{Derivation of the time derivative of the QFI}\label{app:deriv}
The starting point of this derivation is Eq.~\eqref{qfi} for the QFI of a single mode state together with the equations of motion of the first an second statistical moments of the Gaussian state~\eqref{difsig} and~\eqref{difr}.

The time derivative of the QFI~\eqref{qfi} is
\be
\label{eq:QFIder1}
\begin{split}
\dot{I}_{\theta} = & \frac{1}{2} \frac{(1+\mu^2) \Tr [ \partial_t \left( \sig^{-1} \sig \right)^2 ] - \Tr[  \left( \sig^{-1} \sig \right) ] \partial_t \mu^2 }{(1+\mu^2)^2} \\
& + \frac{2 (1 - \mu^4) \partial_t ({\mu'}^2) + 2 \mu'^2 \partial_t \mu^4 }{(1-\mu^4)^2} \\
&+ 2 (\partial_t \rvec')^\T \sig^{-1} \rvec' + 2 \rvec'^\T \sig^{-1} (\partial_t \rvec)
+ 2 \rvec'^\T (\partial_t \sig^{-1}) \rvec'
\end{split}
\ee

Finally, we need the equation of motions for the derivatives of the first moment vector and of the CM, obtained by differentiating~\eqref{difsig} and~\eqref{difr} with respect to the parameter:
\begin{align}
\dot{\sig'}= - \sig' \label{difdsig} \\
\dot{\rvec'}= - \frac{\rvec'}{2} \label{difdr}.
\end{align}

In this calculation we will use the  the following formula for the derivative of the determinant of an invertible matrix $A(t)$: $\frac{\mathrm{d}}{\mathrm{d} t} \mathrm{Det} A(t)  = \mathrm{Det} A(t) \Tr \left[ A(t)^{-1}\frac{\mathrm{d}}{\mathrm{d}t} A(t) \right] $, as well as the formula for the derivative of the inverse matrix $\frac{\mathrm{d}}{\mathrm{d} t} A(t)^{-1}  =- A(t)^{-1} \left( \frac{\mathrm{d}}{\mathrm{d}t} A(t) \right) A(t)^{-1}$.

\subsection*{First term}
We first start by noticing that
\be
\Tr\left[ \partial_t \left(  \sig^{-1} \sig' \right)^2 \right] = 2 \Tr \left[ ( \sig^{-1} \sig') \partial_t ( \sig^{-1} \sig' ) \right] ,
\ee
where we have used the cyclicity of the trace.
This term can be simplified as follows
\be
\begin{split}
& ( \sig^{-1} \sig') \partial_t ( \sig^{-1} \sig' ) =  ( \sig^{-1} \sig') (\partial_t \sig^{-1} \sig' +  \sig^{-1} \partial_t \sig' ) \\
&=  ( \sig^{-1} \sig') (-\sig^{-1} (\partial_t \sig) \sig^{-1} \sig' +  \sig^{-1} \partial_t \sig' ) \\
&= ( \sig^{-1} \sig') (-\sig^{-1} (-\sig + N \id ) \sig^{-1} \sig' +  \sig^{-1} (-\sig') ) \\
& = - N \sig^{-1} \sig' \sig^{-2} \sig' \, ,
\end{split}
\ee
so that we obtain
\be
\label{eq:1t1}
\Tr\left[ \partial_t \left(  \sig^{-1} \sig' \right)^2 \right] = -2 N \Tr \left[ (\sig^{-1}\sig')^2 \sig^{-1}\right] \,.
\ee

The second part of the first term's numerator can be expanded using the following identity
\be
\label{eq:1t2}
\begin{split}
\partial_t \mu^2 &= \partial_t ( \mathrm{Det} \sig )^{-1} = - \frac{1}{(\mathrm{Det} \sig)^2} ( \partial_t \Det \sig ) \\ 
& = - \frac{1}{\mathrm{Det} \sig} \Tr \left[ \sig^{-1} \partial_t \sig \right] \\
& = - \frac{1}{\mathrm{Det} \sig} \Tr \left[ \sig^{-1} (-\sig + N \id ) \right] \\
& = - \mu^2 \left( N \Tr [ \sig^{-1} ] -2 \right) \,.
\end{split}
\ee
Using Eqs.~\eqref{eq:1t1} and~\eqref{eq:1t2}, from the first term in~\eqref{eq:QFIder1} we obtain the first line of Eq.~\eqref{idiot}.

\subsection*{Second term}
First of all, using the definition of the purity $\mu=(\Det\sig)^{-1/2}$ we find that
\be
\mu' = - \frac{1}{2} (\Det\sig)^{-1/2} \Tr \left[ \sig^{-1} \sig' \right] = - \frac{\mu}{2} \Tr \left[ \sig^{-1} \sig' \right]
\ee
and analogously
\be
\partial_t \mu = - \frac{\mu}{2} \Tr \left[ \sig^{-1} \partial_t \sig \right] \,.
\ee

The first term we need to evaluate from the second line of Eq.~\eqref{eq:QFIder1} is the following
\be
\partial_t ({\mu'}^2) = 2 \mu' (\partial_t \mu') = -\mu \Tr \left[ \sig^{-1} \sig' \right] (\partial_t \mu'), 
\ee
where now we need to evaluate this last term 
\be
\begin{split}
\partial_t \mu' &= -\frac{1}{2} (\partial_t \mu) \Tr \left[ \sig^{-1} \sig'\right] -\frac{1}{2} \mu \left( \partial_t \Tr \left[ \sig^{-1} \sig'\right] \right) \\
& = \frac{\mu}{4}  \Tr \left[ \sig^{-1} \partial_t \sig \right] \Tr \left[ \sig^{-1} \sig'\right]  \\
& - \frac{\mu}{2} \left( -\Tr \left[ \sig^{-1} (\partial_t \sig) \sig^{-1} \sig'\right] + \Tr \left[ \sig^{-1} (\partial_t \sig') \right] \right) \\
&=  \frac{\mu}{4}  \Tr \left[ \sig^{-1} (-\sig + N \id ) \right] \Tr \left[ \sig^{-1} \sig'\right]  \\
& - \frac{\mu}{2} \left( -\Tr \left[ \sig^{-1} (-\sig + N \id ) \sig^{-1} \sig'\right] + \Tr \left[ \sig^{-1} (-\sig') \right] \right) \\
&= \frac{\mu}{4} \left\{ \left( N \Tr \left[ \sig^{-1} \right] -2 \right)\Tr \left[ \sig^{-1} \sig'\right] + 2 N \Tr \left[ \sig^{-2} \sig'\right] \right\}
\end{split}
\ee
We thus get to
\be
\label{eq:2t1}
\begin{split}
\partial_t ({\mu'}^2) = -\frac{\mu^2}{4} \Tr \left[ \sig^{-1} \sig' \right] &\biggl\{ \left( N \Tr \left[ \sig^{-1} \right] -2 \right)\Tr \left[ \sig^{-1} \sig'\right] \\ 
& + 2 N \Tr \left[ \sig^{-2} \sig'\right] \biggr\}
\end{split}
\ee

For the next part of the second line of Eq.~\eqref{eq:QFIder1} we need to evaluate the following term 
\be
\label{eq:2t2}
\begin{split}
\mu'^2 \partial_t \mu^4 &= \left(- \frac{\mu}{2} \Tr \left[ \sig^{-1} \sig' \right] \right)^2 4 \mu^3 \left( -\mu \Tr \left[ \sig^{-1} \partial_t \sig \right] \right) \\
& = - \mu^6 \Tr \left[ \sig^{-1} \sig' \right] \left( N \Tr \left[ \sig^{-1} \right] - 2 \right) .
\end{split}
\ee
From Eqs.~\eqref{eq:2t1},~\eqref{eq:2t2} and~\eqref{eq:QFIder1} we obtain the first term on the third line of Eq.~\eqref{idiot}.

\subsection*{Third term}
We consider the three terms on the last line of~\eqref{eq:QFIder1} together and we find
\be
\begin{split}
&2 (\partial_t \rvec')^\T \sig^{-1} \rvec' + 2 \rvec'^\T \sig^{-1} (\partial_t \rvec)
+ 2 \rvec' (\partial_t \sig^{-1}) \rvec' \\
&= - 2 {\rvec'}^\T \sig^{-1} \rvec' - 2 \rvec'^\T \sig^{-1} (\partial_t \sig ) \sig^{-1} \rvec' \\ 
& = - 2 {\rvec'}^\T \sig^{-1} \rvec' - 2 \rvec'^\T \sig^{-1} (-\sig + N \id  ) \sig^{-1} \rvec' \\
& =   - 2 N \rvec'^\T \sig^{-2} \rvec' \; ,
\end{split}
\ee
which is the last term in Eq.~\eqref{idiot}. 

\nocite{apsrev41Control}
\bibliography{GaussianQFIControl}

\end{document}